\documentclass{IOS-Book-Article}

\usepackage{graphicx}
\usepackage{mathptmx}

\usepackage{xspace}

\def\hb{\hbox to 10.7 cm{}}

\newcommand{\exanest}{ExaNeSt\xspace}
\newcommand{\euroexa}{EuroExa\xspace}
\newcommand{\apenet}{APEnet\xspace}

\newcommand{\lofamo}{LO{\textbar}FA{\textbar}MO\xspace}
\newcommand{\quong}{QUonG\xspace}
\newcommand{\ie}{\textit{i.e.}\xspace}
\newcommand{\eg}{\textit{e.g.}\xspace}
\newcommand{\apelink}{APElink\xspace}

\newcommand{\exanet}{ExaNet\xspace}
\newcommand{\exaip}{\exanet Network IP\xspace}

\newcommand{\aperouter}{APErouter\xspace}
\newcommand{\apepacket}{APEpacket\xspace}
\newcommand{\apephy}{APEphy\xspace}

\begin{document}

\begin{center}
The final publication is available at IOS Press through \\ 
http://dx.doi.org/10.3233/978-1-61499-843-3-750 \\
(2018) Advances in Parallel Computing, 32, pp. 750-759, Talk at ParCo 2017.
\end{center}

\pagestyle{headings}
\def\thepage{}

\begin{frontmatter}              

\title{Large scale low power computing system} 
\subtitle{Status of network design in \exanest and \euroexa projects}


\author[A]{\fnms{Roberto}        \snm{Ammendola}}
\author[B]{\fnms{Andrea}         \snm{Biagioni}}
\author[B]{\fnms{Fabrizio}       \snm{Capuani}}
\author[B]{\fnms{Paolo}          \snm{Cretaro}}
\author[B]{\fnms{Giulia}         \snm{De Bonis}}
\author[B]{\fnms{Francesca}      \snm{Lo Cicero}}
\author[B]{\fnms{Alessandro}     \snm{Lonardo}}
\author[B]{\fnms{Michele}        \snm{Martinelli}}
\author[B]{\fnms{Pier Stanislao} \snm{Paolucci}}
\author[B]{\fnms{Elena}          \snm{Pastorelli}}
\author[B]{\fnms{Luca}           \snm{Pontisso}}
\author[B]{\fnms{Francesco}      \snm{Simula}}
and
\author[B]{\fnms{Piero}          \snm{Vicini}
\thanks{Corresponding Author: Piero Vicini, INFN Sezione di Roma, Piazzale Aldo Moro 2, Roma, Italy\\  
E-mail:piero.vicini@roma1.infn.it}}

\address[A]{INFN, Sezione di Roma Tor Vergata, Italy}
\address[B]{INFN, Sezione di Roma, Italy}

\begin{abstract}
The deployment of the next generation computing platform at ExaFlops scale requires to solve new technological challenges mainly related to the impressive number (up to $10^6$) of compute elements required. This impacts on system power consumption, in terms of feasibility and costs, and on system scalability and computing efficiency. In this perspective analysis, exploration and evaluation of technologies characterized by low power, high efficiency and high degree of customization is strongly needed.
Among the various European initiative targeting the design of ExaFlops system, \exanest and \euroexa are EU-H2020 funded initiatives leveraging on high end MPSoC FPGAs. Last generation MPSoC FPGAs can be seen as non-mainstream but powerful HPC Exascale enabling components thanks to the integration of embedded multi-core, ARM-based low power CPUs and a huge number of hardware resources usable to co-design application oriented accelerators and to develop a low latency high bandwidth network architecture.

In this paper we introduce \exanet the FPGA-based, scalable, direct network architecture of \exanest system. \exanet allow us to explore different interconnection topologies, to evaluate advanced routing functions for congestion
control and fault tolerance and to design specific hardware components
for acceleration of collective operations.
After a brief introduction of the motivations and goals
 of \exanest and \euroexa projects, we will
report on the status of network
architecture design and its hardware/software testbed adding preliminary bandwidth and latency
achievements.
\end{abstract}

\end{frontmatter}
\markboth{October 2017\hb}{October 2017\hb}

\section{Introduction}
The next generation HPC systems will be characterized by ExaFlops
performances --- \ie $10^{18}$ Floating Point Operations per second
--- and will be used in a broad class of large size,
\mbox{computing-demanding} scientific and industrial applications,
ranging from modeling and simulation of complex physical systems to
biotechnology, cloud computing, big data and analytics.

At a very high level of abstraction, an \mbox{ExaFlops-scale} HPC
system will be composed of something in the order of millions of
computing cores.
A simple system power estimation, based on current
\mbox{state-of-the-art} technologies and $10^6$ compute elements,
yields a total power consumption of $\sim100$MW which is a factor
5$\div$10 more than what is generally agreed as the maximum possible
for a single data center site.

So, if in the past the challenge was to achieve the floating point
performance target with the minimal number of computing components and
limited procurement costs, today the computing efficiency, the power
consumption and the system cooling have become key factors driving HPC
design.
In this perspective, the next generation \textit{one-million cores}
HPC systems have to adopt low power CPUs, \mbox{co-design} and
integrate \mbox{application-oriented} computing accelerators, build
new cooling mechanics while designing scalable, low latency, high
performance interconnection architectures suitable for such as extreme
scale systems.

In this paper we describe the architecture and the FPGA implementation
of \exanet, a novel network architecture targeted for Exascale HPC
systems which is under design in the framework of H2020
\mbox{EU-funded} projects, \exanest and \euroexa, whose motivations
and goals are described in section \ref{H2020-projects}.
Section~\ref{sec:exanet} details the \exanet architecture and hardware
implementation while section~\ref{sec:karma} describes the KARMA test
framework, discussing power consumption estimations and preliminary
latency and bandwidth test results.

\section{Designing European ExaScale systems:  \exanest and \euroexa H2020 project} 
\label{H2020-projects}

In the last 5 years a number of \mbox{EU-funded} initiatives were
launched, trying to reduce the technological gap between European and
\mbox{extra-European} HPC systems.
Leveraging on ARM architectures, the \mbox{Mont-Blanc} series of
projects~\cite{montblanc:2016:short} has paved the way towards an HPC
system based on low power CPUs while the
UniServer~\cite{Marazakis:EUROSERVER:2016:short} project has designed
a scalable platform based on \mbox{ARM-based} microservers.

The \exanest~\cite{DSD:EXANEST:2016:short} project, started on
December 2015 and funded in EU H2020 research framework (call
\mbox{H2020-FETHPC-2014}, n. 671553), is a pillar of a larger
initiative that includes ExaNode~\cite{EXANODE} and
EcoScale~\cite{Mavroidis:ECOSCALE:2016:short} projects and aims to
demonstrate the efficient usage of low power architectures in Exascale
computing platforms.
ExaNode focuses on delivering a novel, highly \mbox{power-efficient}
compute element for HPC and EcoScale deploys a full software stack
needed to use \mbox{high-end} FPGAs as accelerators for HPC systems.
Furthermore, \exanest combines industrial and academic research
expertise to design the architecture and deploy a fully functional
demonstrator of an innovative \mbox{system-level} interconnect,
distributed NVM (Non-Volatile Memory) storage and advanced cooling
infrastructure for an \mbox{ARM-based} \mbox{ExaFlops-class}
supercomputer.

One of the main goals within \exanest is the design of
\textit{\exanet}: a novel, unified (for data and storage traffic), low
latency, high throughput, \mbox{RDMA-based} interconnect architecture
suitable for extreme scale system.
The project leans on last generation high end SoC (System on Chip)
FPGAs --- the Xilinx Zynq UltraScale+\cite{Zynq_Ultrascale+:site} with
4 ARM \mbox{Cortex-A53} embedded cores running at up to $1.5 GHz$ ---
to integrate thousands of cores into a fully working system prototype.

Starting in September 2017, a new H2020 EU initiative named
\euroexa\cite{EuroExa:site} --- funded under the \mbox{FETHPC-2016}
program (call \mbox{H2020-FETHPC-2016}, n. 754337) --- will build upon
\exanest results to deliver a \mbox{world-class}, \mbox{ARM-based} HPC
platform prototype.
The main deliverable of the project is a \mbox{co-designed} platform
capable of scaling to a peak performance to 400~PFlops in a system
with a peak power envelope of 30~MW achieved through a customized
\mbox{ARM-based} processing unit, the adoption of FPGAs for
\mbox{data-flow} acceleration and the integration at rack level of a
low latency, high throughput \mbox{\exanet-based} network
architecture.

\section{\exanet}
\label{sec:exanet}

\exanet is responsible for data communication at Tier~0/1/2 of the
network interconnect of the \exanest project.
\exanet is the product of a joint collaboration among the Foundation
for Research and Technology (FORTH) in Greece and Istituto Nazionale
di Fisica Nucleare (INFN) in Italy.
The INFN APE Research group, which in the past has designed the
\apenet~\cite{ammendola2011apenet+:short} \mbox{3D-Torus} network
architecture, is responsible for the \exaip that provides switching
and routing features and manages the communication over the High Speed
Serial (HSS) links through different levels of the \exanest
interconnect hierarchy:
\begin{itemize}
\item the \mbox{high-throughput} \mbox{intra-QFDB} level (Tier~0) for
  data transmission among the four FPGAs of the \exanest node;
\item the \mbox{intra-Mezzanine} level (Tier~1) directly connecting
  the network FPGAs of different nodes within the same mezzanine;
\item \mbox{inter-Mezzanine} communication level (Tier~2) managing the
  connectivity of the Mezzanine based on SFP+ connectors and allowing
  for the implementation of a direct network among QFDBs within a
  Chassis.
\end{itemize}
The \exaip mainly consists of two hardware components:
\begin{itemize}
\item the \aperouter, handling the routing and switching mechanism of
  the network IP as described in Section~\ref{sec:aperouter};
\item the \apelink I/O interface, managing the data transfers over the
  HSS links as reported in Section~\ref{sec:apelink}.
\end{itemize}

\subsection{\exanet development platform.}
\label{sec:testbed}

Being the early stages of the \exanest project, the current \exanet
release was designed on a Trenz~\cite{Trenz:site} TEBF0808 system,
featuring the same Xilinx Ultrascale+ MPSoC FPGA family chosen for the
final prototype (XCZU9EG).
Preliminary tests were performed to validate the network, connecting
up to four boards shaping a $2 \times 2$ mesh topology through the two
SFP+ connectors available on each Trenz system.

The testbed allows to validate the adoption of the \apenet
architecture at both Tier~0 and Tier~1.
The QFDB composed by four FPGAs matches perfectly with the testing
platform.
Furthermore, the development platform emulates the communication among
the four network FPGAs of the QFDBs hosted within the \mbox{track-1}
mezzanine.

\subsection{Packet structure}
\label{sec:exapacket}

\begin{figure}[!hbt]
\centering
  \includegraphics[width=.95\textwidth]{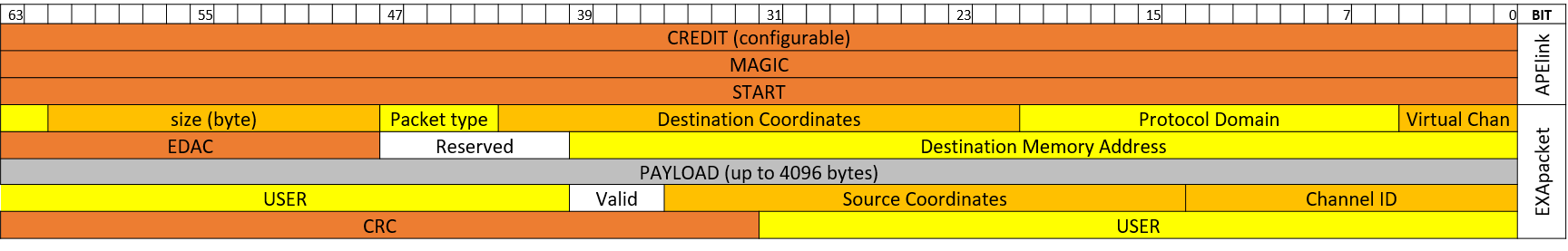}
  \caption{Format examples of packet and \apelink protocol of the
    \exanet interconnect.}
  \label{fig:exapacket}
\end{figure}

A modified version of the \apepacket is the data structure of the
\exanet communication system based on the latest generation of the
\apenet protocol.
Figure~\ref{fig:exapacket} outlines the \exanet packet.

The packet is composed by a \mbox{128-bit} \textit{header}, a
\mbox{128-bit} \textit{footer} and a \textit{payload}.
The maximum payload size is $256\div512$~bytes, being a good
compromise between bandwidth performances and routing efficiency (to
avoid delaying high priority packets).

\subsection{\aperouter}
\label{sec:aperouter}

The \aperouter block dynamically interconnects the intra-tile ports
--- \ie the interface between the programming logic and the
programming subsystem --- and \mbox{inter-tile} ports --- \ie the I/O
interface with the other nodes --- and comprises a fully connected
switch, plus routing and arbitration blocks.

The current release of the \aperouter is targeted to Xilinx devices
and is compliant with the \exanet Header format.
It manages different kinds of packets and supports \mbox{byte-aligned}
data structures in memory --- the previous releases were
\mbox{word-aligned}.
The design has been tuned to evaluate different network topologies and
to explore the introduction of new features --- adaptive routing
algorithms and hardware acceleration of collective functions.

Although the basic functionalities of the \aperouter were verified in
the past --- on the \quong~\cite{ammendola2011quong} prototype located
in Rome --- correct behaviour of the new release is currently under
test on the \mbox{mini-cluster} composed by Trenz boards described in
Section~\ref{sec:testbed}.
Single- and \mbox{multiple-hop} tests were performed and the results
are shown in Section~\ref{sec:karma} and Section~\ref{sec:bw}.

\begin{figure}[!hbt] 
\centering
  \includegraphics[width=0.75\textwidth]{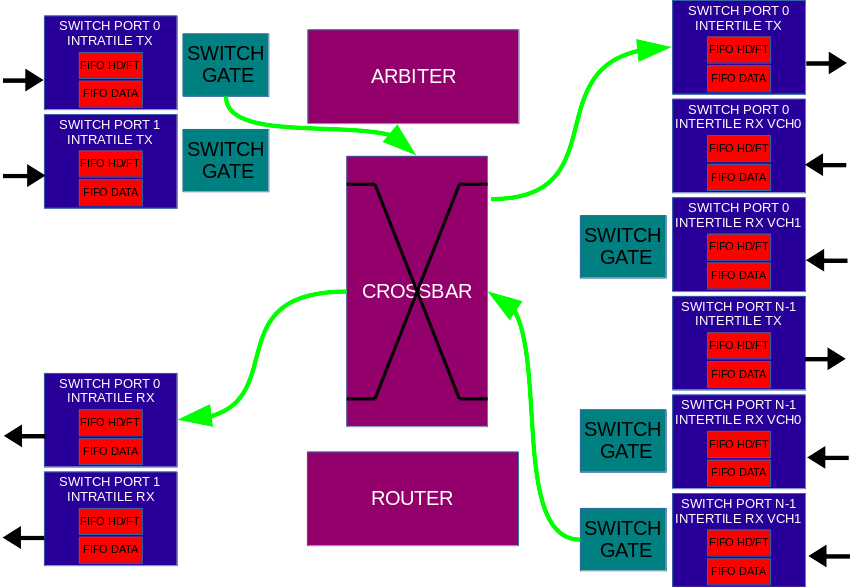}
  \caption{The block diagram of the \aperouter on \exanet prototype.}
\label{fig:aperouter}
\end{figure}

The block diagram of the \aperouter is depicted in
Figure~\ref{fig:aperouter}.

The bidirectional \textbf{Switch Port} contains transmitting (TX) and
receiving (RX) FIFOs.
The \mbox{header/footer} FIFOs are $128 \times 128$~bit (2~KB), while
the \mbox{intra-tile} and \mbox{inter-tile} payload FIFO are $4096
\times 128$~bit (64~KB) and $1024 \times 128$~bit (16~KB)
respectively.
  
The \textbf{Switch Gate} connects data and control signals coming from
the \mbox{intra-tile} and \mbox{inter-tile} ports with the crossbar.
It manages the data flow preventing FIFOs overflow and guarantees
proper transmission of the packet sequence --- header, payload,
footer.

The \textbf{Router} applies a deterministic \mbox{Dimension-Ordered}
Routing (DOR) policy consisting in reducing to zero the offset between
current and destination node coordinate along one dimension before
considering the offset in the next dimension.
The router is able to simultaneously handle more than one packet
transaction.
Specialized priority registers allow selecting the coordinates
evaluation order --- \ie first Z is consumed, then Y and finally X ---
and disabling ports altogether.
The implemented switching technique is Virtual
\mbox{Cut-Through}~\cite{Kermani79virtualcut-through:} (VCT): the
router starts forwarding the packet as soon as the algorithm has
picked a direction and the buffer used to store the packet has enough
space.
The DOR algorithm, not \mbox{per-se} \mbox{deadlock-free}, is made as
such by the implementation of two virtual channels for each physical
channel; the router sends packets using the upper virtual channel if
the offset between current and destination node is greater than zero,
the lower virtual channel otherwise.

The \textbf{Arbiter} manages conflicts among the requests, granting
ordered access to packets coming from different ports that request the
same destination port.
The scheduling algorithm is configurable: Round Robin or Fixed
priority --- the latter can be modified at \mbox{run-time} writing the
proper configuration register.

\subsection{\apelink}
\label{sec:apelink}

Packet payload is encapsulated in a lightweight protocol sketched in
Figure~\ref{fig:exapacket}.
Two words --- \mbox{Magic/Start} --- are included into the data flow
over the serial links to frame the packet boundaries and announce the
transmission of the Header; their transmission takes 2~clock cycles
only.

Since misrouted packets are disruptive for the network, the highly
critical header integrity is protected by an Error Correction Code
(ECC).
Payload integrity is guaranteed by a CRC32 code.
%
\apelink does not provide any acknowledgement or retransmission
mechanism, to not affect performance of the transmission forcing
the implementation of additional memory buffers. 

Buffer availability is measured by credit; exchanging credits by two
communicating nodes is mandatory to avoid buffer overflow. 
Outbound words consume it, causing transmission suspension as soon as a
programmable credit threshold (TRED) is reached –-- \ie credit is
exhausted –-- and resuming as soon as info about newly available space
bounces back to the transmitter –-- \ie credit is eventually
restored. This information is exploited by the router to manage the data
flow implementing the VCT switching mechanism.

Besides, some information regarding the health of the node can be
optionally embedded in the credits, allowing for a fault communication
mechanism --- \lofamo~\cite{LOFAMO:FGCS:2014:short} --- that avoids
single points of failure and guarantees a fast broadcast of critical
status to neighboring nodes.
This embedding of diagnostic messages in the communication protocol
limits the amount of additional overhead (no custom diagnostic packets
are necessary) and prevents this flow from affecting overall
performance.

\subsubsection{\apelink hardware IP}
\label{sec:apelinkhw}
The \apelink hardware IP manages the communication protocol over the
serial links, adapting the \mbox{inter-tile} port interface of the
\aperouter with the outbound interface of the network adapter.
The \apelink IP consists of two main components:
(i) the Transmission Control Logic (TCL), a totally FPGA
\mbox{vendor-independent} IP, that manages data and credit flow over
the link (OSI Data Link, \apelink) and
(ii) the Transceiver, provided by the FPGA vendor and configured to
fit the design requirements, implements the OSI Physical layer,
\apephy).

\begin{figure}[!hbt] 
\centering
  \includegraphics[width=\textwidth]{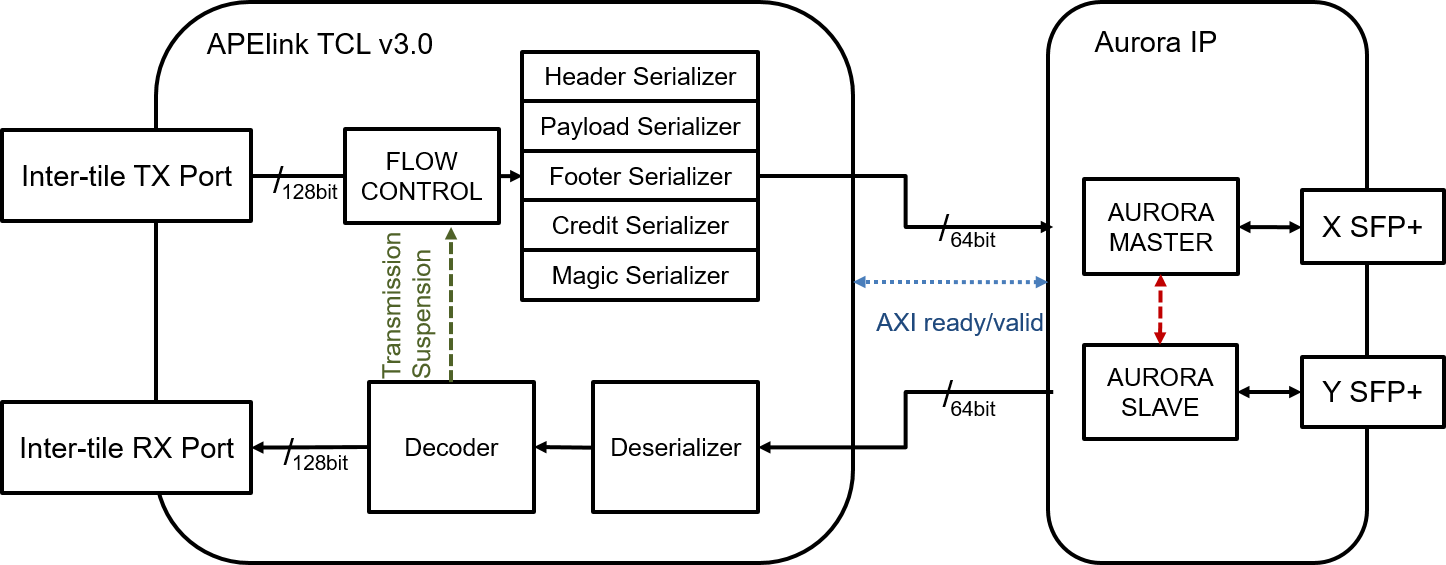}
  \caption{\apelink block scheme.}
\label{fig:apelink}
\end{figure}

The interface between TCL and Transceiver is based on a standard
\mbox{Ready/Valid} mechanism, to be compliant with the AXI stream
protocol and to increase the compatibility of the \apelink hardware
component with different FPGA vendor IPs or custom transceiver
controllers.
The block diagram of current \exanet \apelink data transmission system
is shown in Figure~\ref{fig:apelink}.

The transceiver IP is based on Xilinx Aurora \mbox{64B/66B}
core.
Aurora \mbox{64B/66B} is a lightweight, serial communications protocol
for \mbox{multi-gigabit} links.
It is used to transfer data between devices using one or many GTH
transceivers.
\section{KARMA Test Framework}
\label{sec:karma}

King ARM Architecture (KARMA) is a \mbox{software-oriented} test
framework to validate the \exaip.
The main idea behind its design is the use of the multicore ARM
\mbox{Cortex-A53} Programming System (PS) to emulate in software the
functionalities of the Network Interface (NI), exploiting the AXI low
latency communication capabilities between the PS and the Programming
Logic (PL) that implements the system under test.
This approach turned out to be very effective, allowing for test and
validation of the \exaip since the earliest stages of its development.
It also enabled the rapid prototyping of various architectural
solutions for the interface between the NI and the Switch systems.
Finally, using the framework we were able to characterize the
performance of the two systems in terms of latency.

\begin{figure}[!hbt] 
\centering
  \includegraphics[width=\textwidth]{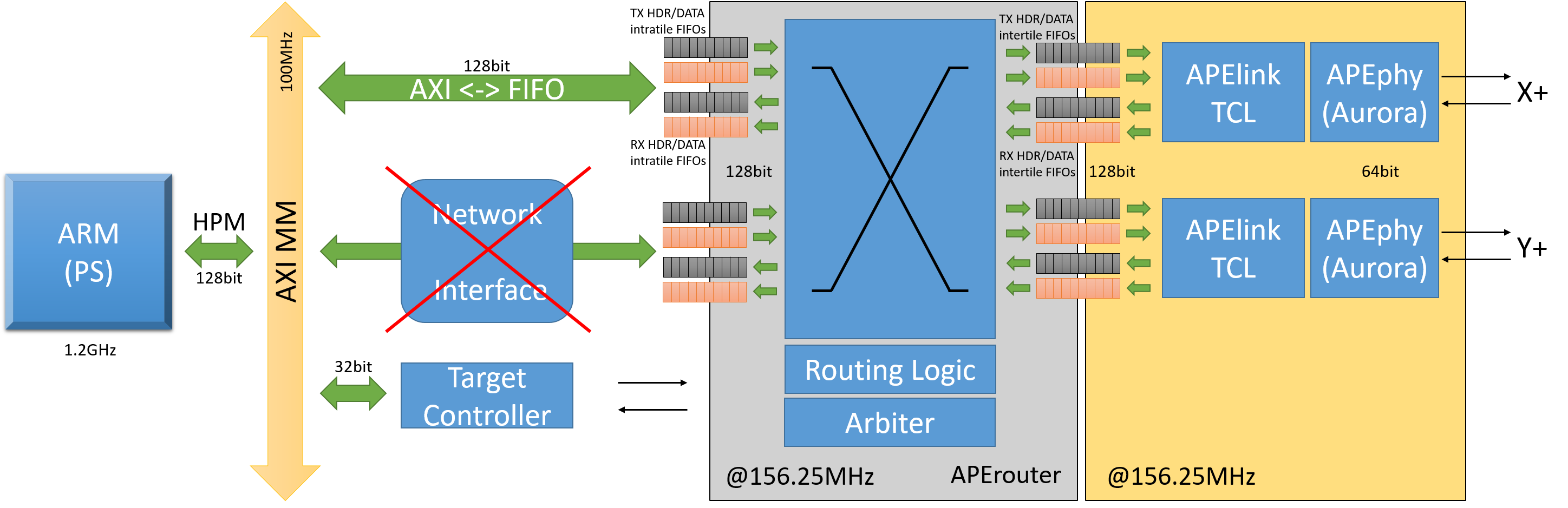}
  \caption{KARMA test framework for the \exaip validation.}
\label{fig:karma}
\end{figure}

On the hardware side, the \mbox{intra-tile} ports are directly
connected to the ARM HPM AXI port through an adapter IP, whose only
purpose is the conversion between streaming and \mbox{memory-mapped}
AXI protocols.

Current KARMA does not implement any \mbox{DMA-access} to the
\mbox{intra-tile} ports, so that ARM must issue a write for every
single word into \mbox{header/data} FIFOs, which is obviously
suboptimal for bandwidth but appropriate for gauging the latency of
\mbox{small-sized} packets.

Moreover, a set of \mbox{configuration/status} registers is accessible
on the same AXI bus through the ``Target Controller'' IP, which allows
configuration of the router (\eg setting coordinates and lattice size)
and probing FIFOs and link status.

An overview of the KARMA test framework is depicted in
Figure~\ref{fig:karma}, where the \exaip --- \ie $3\times2$ \aperouter
and two \apelink Transmission Control Logic blocks and Target
Controller --- is the device under test.
Its operating frequency jointly with the \apephy, composed by the FPGA
embedded transceiver, is 156.25~MHz in order to achieve 10~Gbps on the
serial side.
The Zynq Ultrascale+ Programming Subsystem --- \ie the
\mbox{Quad-core} ARM Cortex-A53 --- works at a frequency of 1.5~Ghz.

The resource usage of the \exaip is reported in
Table~\ref{tab:karmares}.

\begin{table}[!hbt]
\footnotesize
\centering
\begin{tabular}{|c|c|c|c|c|c|c|}
\hline
\textbf{IP}                     & \textbf{LUT}    & \textbf{LUT FF} & \textbf{Registers} & \textbf{BRAM}     & \textbf{GTH}  \\
\hline
\textbf{\exaip}                 & 17287 ($6.3\%$) & 5577 ($6.3\%$)  & 18954 ($6.3\%$)     & 116.5 ($12.7\%$) & 0             \\
\aperouter $3\times2$           & 9599  ($3.5\%$) & 3162 ($1.2\%$)  & 7649  ($1.2\%$)     & 0                & 0             \\
\apelink TCL (2x)               & 5253  ($2.0\%$) & 1698 ($0.6\%$)  & 4854  ($0.8\%$)     & 0                & 0             \\
Target Controller               & 2468  ($0.9\%$) & 187  ($0.1\%$)  & 6451  ($1.1\%$)     & 0                & 0             \\
\hline
\textbf{\apephy Aurora}         & 829   ($0.3\%$) & 488  ($0.2\%$)  & 3107  ($0.6\%$)     & 0                & 2  ($12.5\%$) \\
\hline
\end{tabular}
\caption{KARMA hardware resources overview.}
\label{tab:karmares}
\end{table}

\subsection{Power Consumption}
\label{sec:power}

\begin{figure}[!hbt] 
\centering
  \begin{minipage}[t]{.48\textwidth}
    \includegraphics[width=\textwidth]{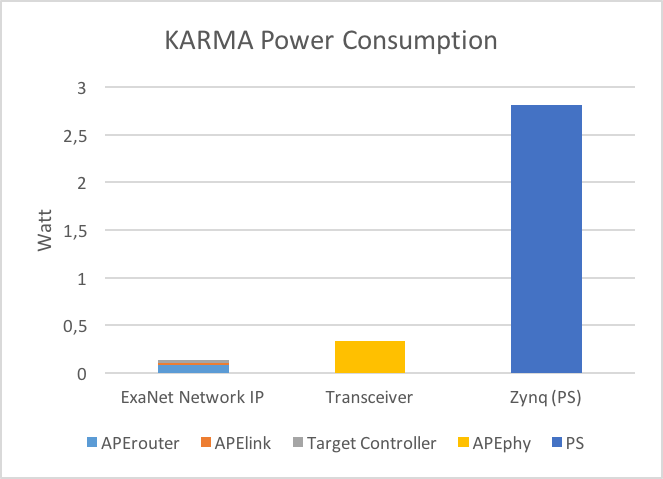}
    \caption{\exaip power consumption.}
    \label{fig:exaippower}
  \end{minipage}
  \quad
  \begin{minipage}[t]{.48\textwidth}  
    \centering
    \includegraphics[width=\textwidth]{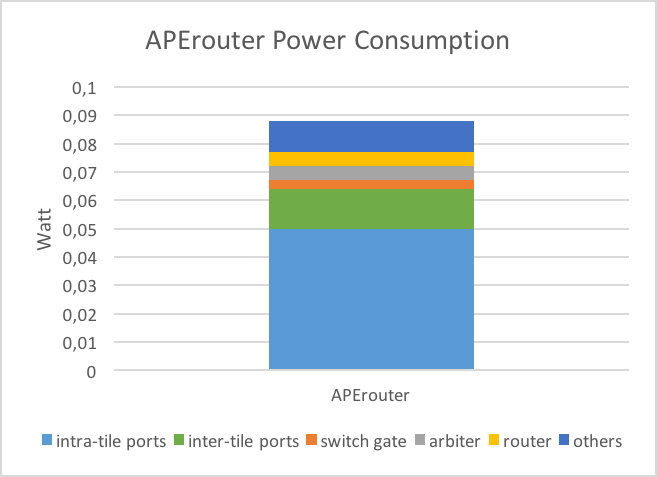}
    \caption{\aperouter power consumption.}
    \label{fig:aperouterpower}
  \end{minipage}
\end{figure}

The Figure~\ref{fig:exaippower} shows the power consumption as
estimated by the Xilinx tool for the KARMA platform.

The \exaip drains 0.136~W only.
The result is strongly dependent by the number of intra- and
\mbox{inter-tile} ports provided as shown in
Figure~\ref{fig:aperouterpower}.
 
The total \aperouter power consumption is 0.088~W, whereas 72\% is the
fraction taken by the inter- and \mbox{intra-tile} ports.
Thus, the size of the FIFOs and the number of implemented virtual
channels modify the final result.
Although a fine tuning should reduce the values, the achieved result
is encouraging nonetheless.
The power consumption for each \apelink TCL is negligible (0.009~W)
while the Aurora transceivers of the \apephy consumes 0.337~W.
The amount of channel provided by the I/O interface is indeed the main
factor.
Finally, the Zynq Ultrascale+ drains 2.822~W, thus the total power
consumption of each board of the development platform is 3.5~W.

\subsection{Latency test}
\label{sec:karmalat}

\begin{figure}[!hbt]
\centering
  \begin{minipage}[t]{.48\textwidth}
\centering
    \includegraphics[width=.95\textwidth]{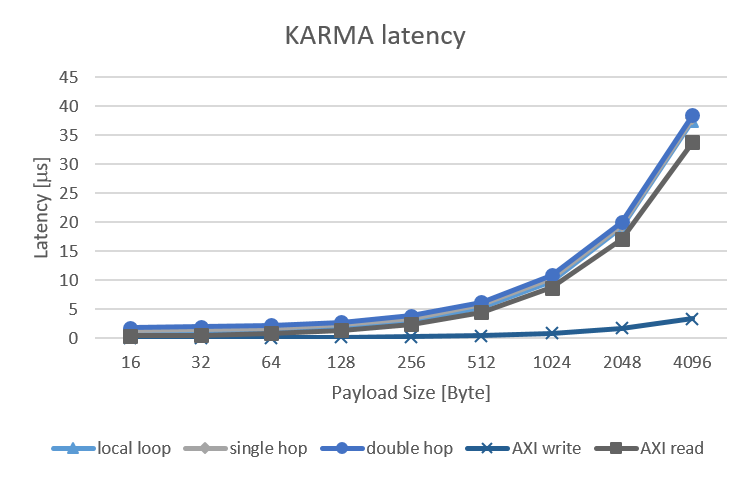}
    \caption{The Roundtrip latency for one and two hops.}
    \label{fig:tlat}
  \end{minipage}
  \quad
  \begin{minipage}[t]{.48\textwidth}
\centering
    \includegraphics[width=.95\textwidth]{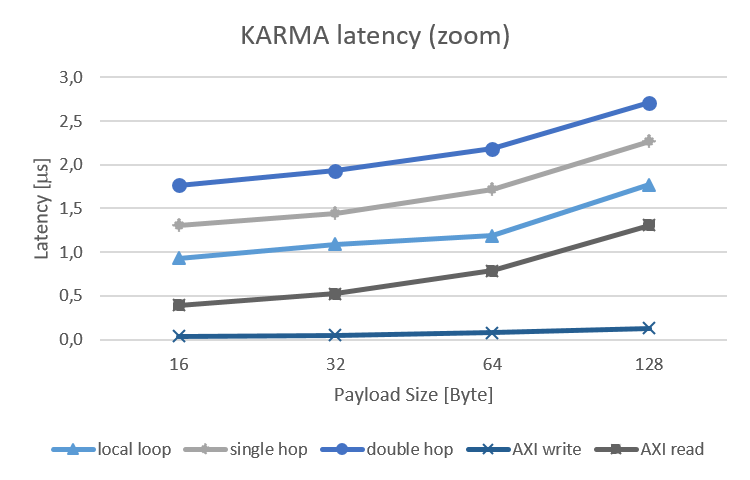}
    \caption {A \mbox{small-packet}, up to 128~Byte --- zoom of the roundtrip latency.}
    \label{fig:tlatzoom}
  \end{minipage}
\end{figure}

The measurements of the \mbox{round-trip} latencies between two boards
--- shown in Figure~\ref{fig:tlat} and Figure~\ref{fig:tlatzoom} ---
were taken bypassing the kernel driver to avoid the notoriously
\mbox{non-optimal} \mbox{bounce-buffering} mechanism and the slow
interrupt handling by \mbox{GNU/Linux}.
This was done implementing a \mbox{user-space} \mbox{ping-pong}
application which exploits \texttt{/dev/mem} to directly access the
\mbox{memory-mapped} hardware.

The stated difference in time of $0.46\mu$s for the two- and
\mbox{one-hop} measurements provides an estimate of the
\mbox{single-hop} traversal time contribution to the total latency.
The times spent by the ARM in reading ($\sim0.4\mu$s, about 20~clock
cycles per word) and in writing ($<0.1\mu$s, 4~clock cycles per word)
on the \mbox{intra-tile} port are independent from the number of hops.

\subsection{Hardware Bandwidth Test}
\label{sec:bw}

As stated before, the KARMA testbed was not designed with the purpose
of evaluating the bandwidth of the device under test.
Nevertheless, the FPGA firmware provides a \mbox{self-test} mechanism
to measure the bandwidth of the \exaip.

The \mbox{self-test} mechanism is composed by three simple IPs:
(i) the \textit{Traffic Generator} generates EXApackets and fills in
the transmitting FIFOs;
(ii) the \textit{Consumer} flushes the receiving FIFOs avoiding the
overflow;
(iii) the \textit{Performance Counter} samples and stores the clock
cycles needed to complete the data transfers.
Configuration registers can customize these \mbox{self-test} packets
in their type, size, destination coordinates and ports.

\begin{figure}[!hbt]
\centering
  \begin{minipage}[t]{.48\textwidth}
  \centering
    \includegraphics[trim=0mm 0mm 0mm 0mm,clip,width=\textwidth]{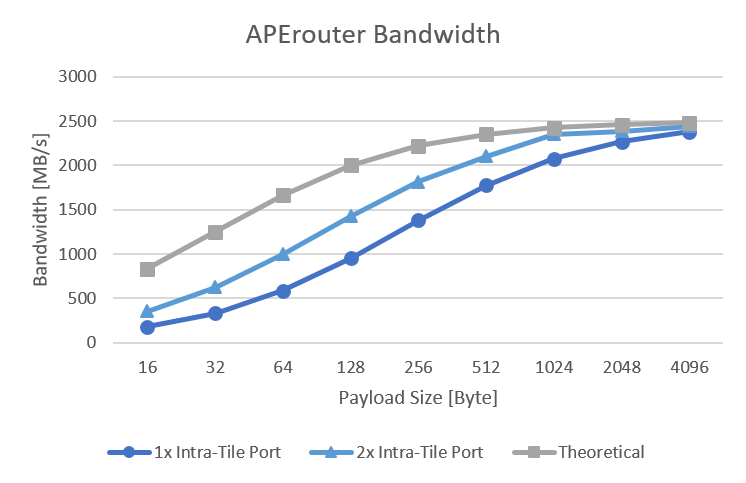}
    \caption {\aperouter bandwidth.}
    \label{fig:aperouterbw}
  \end{minipage}
  \quad
  \centering
  \begin{minipage}[t]{.48\textwidth}
  \centering
    \includegraphics[trim=0mm 0mm 0mm 0mm,clip,width=\textwidth]{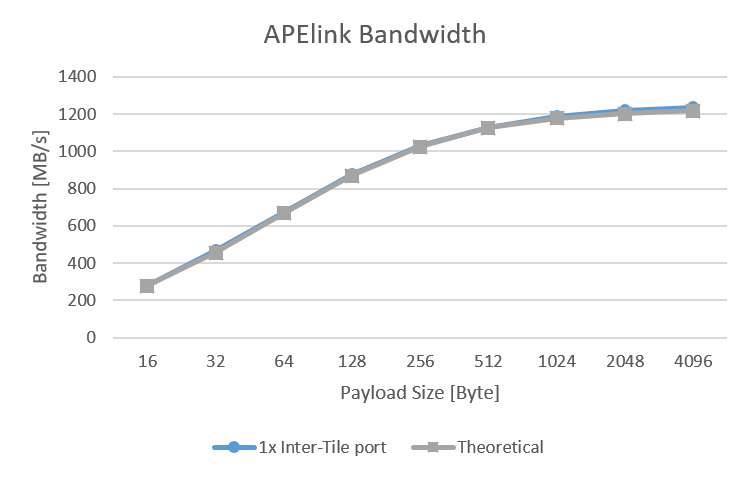}
    \caption {\apelink bandwidth.}
    \label{fig:apelinkbw}
  \end{minipage}
\end{figure}

Figure~\ref{fig:aperouterbw} shows the bandwidth achieved by the
\aperouter when moving data between two ports.
The square markers denote the theoretical peak bandwidth considering
the \mbox{128-bit} bus operating at 156.25~MHz.
The efficiency is $76\%$ for a \mbox{512-byte} packet --- \ie the
maximum packet size --- when the protocol overhead is $6.25\%$.
The performance loss is due to the unoptimized pipeline of the
\aperouter hardware IP.
Some improvements are gained doubling the sending ports (\ie $2\times$
\mbox{Intra-Tile} ports) and transmitting packets to the same target
port; in this case, the efficiency at 512~byte is $89.5\%$.

The \apelink result is shown in Figure~\ref{fig:apelinkbw}.
The theoretical bandwidth is capped at 10~Gbps due to the SFP+
connectors of the Trenz Boards.
The efficiency is $90\%$ for \mbox{512-byte} packets, which is aligned
with the estimate.

\subsection{Conclusion}
In this paper we introduced \exanet: a modular, low latency, high
throughput network architecture suitable for next generation HPC
systems under design in \exanest and \euroexa H2020 projects.
The preliminary results are very encouraging, taking into account the
available margins we have for optimization and improvement:
i) a sub-$\mu S$ \mbox{single-hop}, \mbox{node-to-node} roundtrip
packet transfer latency and
ii) a measured bandwidth close to the peak limit for
\mbox{small-to-medium} size packets.

The next step foresees the improvement of the current \exanet IP to be
integrated in the \exanest final prototype.
The design will focus on adding more \mbox{inter-node} \apelink
channels and \mbox{intra-tile} ports to implement and test \exanest
direct network topologies and sustain the QFDB generated traffic.
The \exanet architecture will be enriched by the addition of new IP
blocks in order to accelerate collective operations and support
adaptive routing to improve network performances and fault tolerance.
In the course of 2018, the \exanet IP will be deployed in the \exanest
final prototype and ported on the coming \euroexa testbeds.

\section*{Acknowledgment}

This work has received funding from the European Union’s Horizon 
2020 Research and Innovation Programme under No. 671553 (ExaNeSt)
and Grant Agreement No. 720270 (HBP SGA1).


\bibliographystyle{iopart-num}
\bibliography{../../../ape_bib/bibliography}

\end{document}